\documentstyle[prb,aps,tighten,preprint,epsf]{revtex}
\ifpreprintsty
\else
\newcommand{\startlongequation}{
\end{multicols}\vspace*{-3.5ex}{\tiny\noindent
\begin{tabular}[t]{c|} \parbox{0.493\hsize}{~} \\ \hline \end{tabular}} }

\newcommand{\stoplongequation}{
{\tiny\hspace*{\fill}
\begin{tabular}[t]{|c}\hline\parbox{0.49\hsize}{~} \\ \end{tabular}}
\vspace*{-2.5ex}\begin{multicols}{2} }

\fi

\begin{document}
\widetext

\draft

\title{Fixed-Node Monte Carlo Calculations for the 1d Kondo Lattice
  Model} 

\author{H.J.M. van Bemmel, W. van Saarloos and D.F.B. ten Haaf}
\address{Institute Lorentz, Leiden University,
 P. O. Box 9506, 2300 RA Leiden, The Netherlands}

\date{\today}

\maketitle

\begin{abstract}
  The effectiveness of the recently developed Fixed-Node Quantum Monte
  Carlo method for lattice fermions, developed by van Leeuwen and
  co-workers,   is tested by applying it to the
  1$d$ 
  Kondo lattice, an example of a one-dimensional model with a
  sign problem. The principles of this method and its implementation
  for the Kondo Lattice Model are discussed in detail.  We compare the
  fixed-node upper bound for the ground state energy at half filling
  with exact-diagonalization results from the literature, and
  determine several spin correlation functions. Our `best estimates'
  for the ground state correlation functions do not depend sensitively
  on the input trial wave function of the fixed-node projection, and
  are reasonably close to the exact values. We also calculate the spin
  gap of the model with the Fixed-Node Monte Carlo method. For this it
  is necessary to use a many-Slater-determinant trial state.  The
  lowest-energy spin excitation is a running spin soliton with wave
  number $\pi$, in agreement with earlier calculations.
\vspace*{0.7cm}

{\em dedicated to Hans van Leeuwen at the occasion of his $65^{th}$
  birthday}
\vspace*{0.7cm}
 
\end{abstract}

\pacs{PACS numbers: 71.27.+a,71.10.+x, 75.10.Jm,71.20.Ad }


\newpage

\section{Introduction}\label{introduction}
As is well known, quantum Monte Carlo simulations are plagued by the
so-called sign problem\cite{raedt,suzuki}. The sign problem refers to the fact that when
physical properties are sampled in configurations space, one collects
large positive and negative contributions due to the fact that a
fermion wavefunction is of different sign in different regions of
configuration space. These contributions of opposite sign tend to
cancel, giving results that may be exponentially smaller than the
separate positive and negative contributions. Though the sign problem
can be circumvented in special cases, e.g., for the Hubbard model at
half filling \cite{hirsch}, no general solution has emerged yet from
the various approaches that have been explored to cure it
\cite{boninsegni,loh,sorella,fahy,gubernatis,an,Bemmel}.

In 1990, when B. J. Alder was Lorentz Professor in Leiden, Hans van
Leeuwen became acquinted with the Fixed Node Monte Carlo (FNMC) method
of Ceperley and Alder \cite{ceperley1,ceperley2,ceperley3}, which
avoids the sign 
problem in the context of continuum Green's function Monte Carlo. This
stimulated him to explore the possibility of formulating a lattice
version of FNMC, first with a postdoc, An\cite{an}, and later in
collaborations with the present authors\cite{Bemmel,haaf}. The
formulation of the approach which was developed later\cite{Bemmel} was
shown to be variational\cite{haaf}, i.e. to give an upper bound to the
exact ground state, and is the subject of this paper, which we
dedicate to Hans van Leeuwen. We test this FNMC method for lattice fermions\cite{Bemmel,haaf,an2} on a
simple one-dimensional (1$d$) model for which various results are available, the 1$d$ Kondo
lattice model (KLM) at half filling. This FNMC method involves an
approximation that removes the sign problem in the context of Green's
function Monte Carlo.  Different Monte Carlo techniques that have been
applied to the 1$d$ KLM include the world-line algorithm\cite{Troyer}, a
finite temperature grand-canonical method involving a
Hubbard-Stratonovich transformation\cite{Fye} and the ground state
method developed by Sorella {\em et al.}\cite{Sorella2,Otsuka}.  All
three Monte Carlo methods suffer from the sign problem, even in 1$d$.

The KLM is one of the basic models for correlated fermions.  It can be
obtained as the strong-coupling limit of the periodic Anderson model,
which aims at capturing the essential physics of heavy-fermion
materials\cite{heavy,hewson}.  In the limit of strong on-site repulsion among
the $f$-electrons, a picture emerges of localized $f$-electrons
interacting with a conduction band.  In recent years, the model has
been studied by a variety of methods, including variational
approximations, exact diagonalizations and the density matrix
renormalization group method
\cite{Lacroix,shiba,ferro1,ferro2,para,Ueda,Tsunetsugu,Schlottmann,Wang,Yu}.
This, together with the fact that quantum Monte Carlo simulations of
this model do have a sign problem, makes the 1$d$ KLM a suitable testing
ground for our lattice FNMC 
method.

The lattice version of FNMC gives, like the continuum
version\cite{ceperley1,ceperley2,ceperley3,ceperley4} which inspired
it, upper bounds for the ground state energy\cite{haaf,misnomer}.  It improves
upon a trial wave function for a given Hamiltonian by employing a
Green's function projection method with a modified Hamiltonian in
which all terms which would lead to unwanted sign changes in the
sampling, are treated in a special way. The sign structure of the
resulting approximate wave function, which is the ground state wave
function of the modified Hamiltonian, is the same as that of the
original trial wave.  One obvious immediate question of interest is
how close the FNMC energy estimate is to the exact ground state energy
of the original Hamiltonian. We will study this question by applying
the fixed-node projection to a  trial wave function with a free
 parameter. As we shall see,
for the KLM, the ground state energy obtained in the FNMC at half
filling is quite independent of the precise trial wave, and quite
close to the values obtained in exact diagonalisation. We also compare
the FNMC results for some correlation functions (which do not obey
bounds), with exact results\cite{Ueda,Otsuka,note1} for chains of 
six sites, with coupling constant $J$ equal to $0.2$ and $1.0$. Here
too we find that our best FNMC simulation estimates are rather
independent of the starting trial wave function, and reasonably close
to the exact values. We finally also show how the spin gap of the 1$d$
KLM can be determined with our FNMC, although this is computationally
much more demanding, since a trial wave consisting of a sum of slater
determinants must be used. Good agreement is found with earlier
results\cite{Wang,Yu} in this case.

Before presenting our results, we first briefly discuss the 1$d$
KLM and the reason why sampling it with unrestricted random walks
leads to the sign problem. Then, in section \ref{FNMC}, the principles
of the lattice FNMC are summarized, followed by details of the
implementation for the KLM.  Section \ref{results} gives the
comparison with exact results for small lattices.  In section
\ref{soliton}, our results for the running spin triplet excitation are
discussed.

\section{The Kondo Lattice Model}\label{inKLM}

The Kondo lattice Hamiltonian is given by
\begin{equation} \label{klm}
  {\cal H}_{KLM}=-t\sum_{\langle ij \rangle} \left(
  c^{\dag}_{i\sigma}c_{j\sigma}+ c^{\dag}_{j\sigma}c_{i\sigma}
\right)-\mu\sum_{i}n_{ic}+J\sum_{i}\vec{S}_{if}\cdot\vec{S}_{ic}~.
\end{equation}
The two kinds of electrons, denoted by $c$ for the conduction band and
$f$ for localized levels, have a spin-spin interaction. For the $f$'s,
the constraint is that there has to be precisely one $f$-electron on
every site. In (\ref{klm}) and below, a summation convection is used
for repeated spin indices.

We seek to write the Hamiltonian in a form that is convenient for GFMC
calculations.  If one would treat the $f$'s as spins, which are not
antisymmetrized but form a dynamical background for the conduction
electrons, the total number of up-spin (and of down-spin) {\em
  fermions} would not be conserved by the Hamiltonian. It is not
convenient to use this representation in a GFMC calculation whose
starting trial wave is a fully fermionic mean-field type wavefunction.
If we use the constraint of one $f$-electron per site and the
identities
\begin{equation}
 \vec{S}_{if}=\frac{1}{2}f^{\dag}_{i\sigma}\vec{\tau}_{\sigma\sigma^{\prime}}
  f_{i\sigma^{\prime}}~,~~~
  \vec{S}_{ic}=\frac{1}{2}c^{\dag}_{i\sigma}\vec{\tau}_{\sigma\sigma^{\prime}}
  c_{i\sigma^{\prime}},
\end{equation}
where the components of $\vec{\tau}_{\sigma\sigma^{\prime}}$ denote
the three Pauli matrices, the Hamiltonian can be written in the form
\begin{equation}
  {\cal H}_{KLM}=-t\sum_{\langle ij \rangle} \left(
  c^{\dag}_{i\sigma}c_{j\sigma}+c^{\dag}_{j\sigma}c_{i\sigma}
\right)-\frac{J}{2}\sum_{i}\left( c^{\dag}_{i\sigma}f_{i\sigma}
\right)\left(f^{\dag}_{i\sigma^{\prime}}c_{i\sigma^{\prime}}\right)-\mu^{\prime}\sum_{i}n_{ic}~,
\label{Ham}
\end{equation}
with $\mu^{\prime} \!=\! \mu \!-\! J/4 $.  This is now fully in
fermionic language.  Therefore, Slater determinants of fixed dimension
can be used.
\begin{figure}
\epsfysize=4.5cm
\epsffile{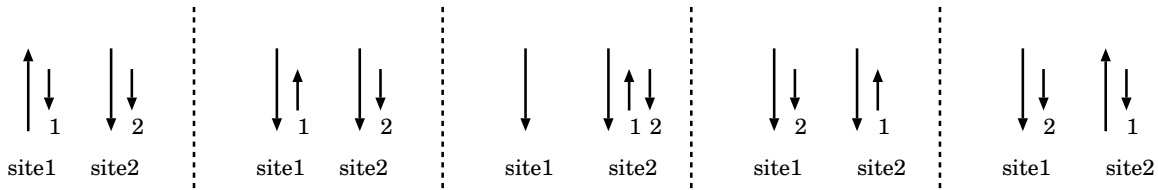}

\vspace*{1.5cm}

\caption[]{\footnotesize A sequence of processes that lets two
conduction electrons of equal spin pass each other in 1$d$.
Large arrows denote $f$-electrons, small ones denote
$c$-electrons. The successive states in the sequence are separated by
vertical dashed lines.}
\end{figure}

In many fermionic lattice models, e.g. the Hubbard model, the fermion
statistics is not really important in Monte Carlo simulations in 1$d$.
The reason is that, fermions of the same spin cannot pass each other
in 1$d$ and so their ordering is fixed. Since the exchanges that give
rise to sign changes as a result of the fermion antisymmetry of the
wave function, are suppressed, there is no sign problem. In higher
dimensions, this is not the case.

For the 1$d$ KLM, there is no fixed ordering of conduction electrons
of equal spin. The presence of the spin flip term proportional to $ J$
in ${\cal H}_{KLM}$ makes it possible for the ordering to change. This
is illustrated in Fig.\ 1. A down $c$-electron at site 1 first changes
in an up-electron, due to the simultaneous flip of a $c$-$f$ pair.
Then, the $c$-electrons at site 1 and 2 have opposite spins, so after
a hop due to the kinetic term, they can occupy the same site. After an
additional hop of the other $c$-electron, the two $c$-electrons then
effectively have been interchanged.

Such an interchange can also happen in a Monte Carlo simulation, and
one needs to take into account that two configurations that differ by
the interchange of two numbered electrons must have opposite signs in
the wave function.  This is the reason why even the 1$d$ KLM exhibits
a sign problem\cite{Troyer,Otsuka}.  In the case of 6 sites at
half-filling with $J\!=\!1.0$, which is studied by
Otsuka~\cite{Otsuka}, it appears that the sign problem is not very
severe, but at certain filling fractions the sign problem is known to
make simulations prohibitively difficult\cite{Troyer}.

The 1$d$ KLM has been studied in different regimes. If the number of
$f$-electrons is equal to the number of sites, and the carrier
concentration is low, there is a ferromagnetic
state\cite{ferro1,ferro2}.  In weak coupling, at larger density but
below half-filling, one obtains a paramagnetic state\cite{para}, and
at half-filling, the system shows insulating spin-liquid
behavior\cite{Fye,Tsunetsugu}. Recently, the ground state was proven
to bea spin singlet and proven to be unique \cite{tsunetsugu2}.  
For slightly less than one $f$-electron per
site, impurity bands arise\cite{Schlottmann}.

In this paper, we limit ourselves to the case of half-filling, one
$c$-electron per site.  Finite-size scaling results\cite{Tsunetsugu}
as well as recent density matrix renormalization group
calculations\cite{Yu} both show that there is a gap for spin and
charge excitations for all $J>0$ and thus confirm the insulating
spin-liquid character of the ground state at half filling.

\section{The FNMC method for the KLM}\label{FNMC}
\subsection{Principles of the FNMC method}
Since the general principles of the FNMC for lattice fermions have
been laid out before\cite{Bemmel,haaf,an},  we only summarize the
most essential aspects of the method here.

In a Green's function Monte Carlo method, one projects out the ground
state of a system with Hamiltonian ${\cal H}$ from an initial trial
state $|\psi_T \rangle$. As before\cite{Bemmel,haaf,an}, we use a
projection operator ${\cal F}$ which acts as follows:
\begin{equation}
  |\psi^{n}\rangle={\cal
    F}^{n}|\psi_{\text{T}}\rangle=\left[1-\tau\left({\cal
    H}-w\right)\right]^{n}|\psi_{\text{T}}\rangle .\label{projector}
\end{equation}
The implementation is in configuration space, and has a stochastic
character. A specific configuration in configuration space, which
determines the locations, spins, etc. of all the labeled electrons, is
denoted by $R$, and we write $\psi_T(R)\!=\!\langle R|\psi_T\rangle$, etc. When
$\tau$ is small enough and $w$ adjusted properly during the sampling
process\cite{an,Trivedi}, the operator ${\cal F}^n$ projects onto the
ground state as $n\!\rightarrow\!\infty$.  To obtain better
statistics, we introduce {\em importance sampling}: we let the Green's
function
\begin{equation}
  G(R,R^\prime )=\psi_{\text{T}}(R)F(R,R^\prime
  )\psi_{\text{T}}^{-1}(R^\prime )
\label{Green}
\end{equation}
determine the transition probabilities of a random walker from
$R^\prime$ to $R$; for simplicity, we take the trial wave function to
be real. The projection (\ref{projector}) then becomes
\begin{equation} \label{prod}
  \psi^n(R) = \sum_{R_n \cdots R_1} \psi_T(R)
  G(R,R_{n})G(R_{n},R_{n-1}) \cdots G(R_3,R_2)G(R_2,R_1) \psi_T^2(R_1)
\end{equation}
In the random walk interpretation underlying the Monte Carlo process,
the initial distribution of the random walkers is given by
$\psi_T^2(R)$, and (\ref{prod}) is sampled stochastically by splitting
$G$ as
\begin{equation}\label{gsplit}
  G(R,R')=P(R,R')m(R')~,
\end{equation}
with $ m(R')\!\equiv\!\sum_{R}G(R,R')$ and hence $\sum_{R}P(R,R')=1$.
This notation anticipates that we wish to view $P$ as a transition
probability, the probability for a particle to make a transition from
configuration $R'$ to $R$, so that a path $R_1,R_2,\cdots,R_n$ in
configuration space is generated by sampling the transitions according
to $P$. The weight factors $m$ which are accumulated along a path are
sampled by viewing them as a multiplicity factor of each
walker\cite{an,Trivedi}.  After a suitable number of steps, these
multiplicity factors are sampled by a branching process: at these
events, a walker with a multiplicity factor $m$ can be either killed,
stay alive, or split into more walkers in such a way that, on average,
there are $<\!m\!>$ new ones after the event. After each branching
event, the factors $m$ of all the walkers are reset to 1.

If all transition probabilities $G$ would be positive, the above
process could be implemented straightforwardly as in simulations of
boson lattice models\cite{Trivedi}. For fermions, the sign problem
arises in the present formulation through the fact that $G(R,R')$, and
hence $P(R,R')$, can be negative. In particular, for the KLM
(\ref{Ham}), as for the Hubbard model\cite{Bemmel}, all off-diagonal
terms $\langle R|H_{KLM}|R'\rangle$ of the Hamiltonian ${\cal
  H}_{KLM}$ are negative, and so negative signs arise in making
transitions between configurations at which the trial wave function
has opposite signs, $\psi_T(R)/ \psi_T(R')\!<\!0$.  Taking those
transition probabilities $P(R,R^\prime )$ proportional to $\left|
G(R,R^\prime )\right| $ and carrying the sign with each walker would
give positive and negative {\em multiplicities}, eventually, and
therefore, the implementation of Eq.(\ref{prod}) would lead to large
positive and negative contributions to all measured quantities. The
resulting cancellations are the essence of the sign problem.

In the lattice FNMC the sign problem is avoided by introducing a
slightly different {\em effective Hamiltonian} ${\cal H}_{eff}$ in
which all steps that lead to a sign change of the walker are left out,
and replaced by a potential term\cite{Bemmel,haaf}. The steps that are
left out satisfy
\begin{equation}
  \langle
  R|H|R^\prime\rangle\psi_{\text{T}}(R)\psi_{\text{T}}(R^\prime) > 0 .
\label{condition}
\end{equation}
This is implemented by defining an effective Hamiltonian as follows:
The off-diagonal terms are
\begin{eqnarray}
  \langle R|H_{\text{eff}}|R^\prime\rangle & \equiv & \langle
  R|H|R^\prime\rangle \hspace*{0.5em} \mbox{(if $\langle
    R|H|R^\prime\rangle\psi_{\text{T}}(R)\psi_{\text{T}}(R^\prime) <
    0$)}~, \nonumber
\\
& \equiv & 0 \hspace*{4.1em} \mbox{(otherwise),} \label{Heffoffdiag}
\end{eqnarray}
and the diagonal terms are
\begin{equation}\label{Heffdiag}
  \langle R|H_{\text{eff}}|R\rangle \equiv \langle R|H|R\rangle +
  \langle R|V_{\text{sf}}|R\rangle ,
\end{equation}
where the `sign flip' potential that replaces the hops that satisfy
Eq.(\ref{condition}) is given by
\begin{equation}\label{Vsf}
  \langle R|V_{\text{sf}}|R\rangle \equiv
  \sum_{R^\prime}^{\text{sf}}\langle
  R|H|R^\prime\rangle\frac{\psi_{\text{T}}(R^\prime)}{\psi_{\text{T}}(R)}
  .
\label{prescription}
\end{equation}
In this expression, the sum runs over all configurations $R'$
connected by a non-zero matrix element $\langle R'|H|R \rangle$ to the
configurations $R$,  for which (\ref{condition}) holds.  The ground
state energy of ${\cal H}_{eff}$, which can be sampled without sign
problem, gives\cite{haaf} an upper bound to the ground state energy of
the true Hamiltonian $\cal H$. Expectation values of physical
quantities are then obtained in the standard way\cite{Trivedi,an}.

\subsection{The variational mean-field type trial  state for the KLM}

As we saw above, a prerequisite for a Green's function Monte Carlo
calculation is a trial wave function.  For the KLM, we use what
amounts to a Gutzwiller-projected mean-field type wavefunction as a
trial state. This wavefunction is essentially obtained as follows.
Since earlier work indicates that this gives the lowest energy
results, we use the Kondo decoupling scheme in Eq.\ (\ref{Ham}), i.e.,
${\cal H}_{KLM}$ is approximated by
\begin{equation}
\label{Hamv}
{\cal H}_{V}=-t\sum_{\langle ij \rangle} \left(
c^{\dag}_{i\sigma}c_{j\sigma}+c^{\dag}_{j\sigma}c_{i\sigma}
\right)-\sum_{i} \left[ \left( c^{\dag}_{i\sigma}f_{i\sigma} \right)
V_i +
\left(f^{\dag}_{i\sigma^{\prime}}c_{i\sigma^{\prime}}\right)V_i^*
\right]+ \frac{4}{J} \sum_{i}| V_i|^2.
\end{equation}
The Hamiltonian ${\cal H}_V$ is bilinear in the fermion operators and
hence can be diagonalized. We denote the ground state of this
Hamiltonian by $| \psi_V \rangle$. The trial state $|\psi_T \rangle$
we use for our calculations is then
\begin{equation}
  | \psi_T\rangle = P_G |\psi_V \rangle,
\label{psiT}
\end{equation}
where $P_G$ is the Gutzwiller projection operator which projects onto
states in which each site is occupied by one
$f$-electron\cite{shiba,Otsuka}.

For the homogeneous ground state, all $V_i$'s should be taken equal,
$V_i=V$.  Following Otsuka\cite{Otsuka}, we will use this $V$ as a
variational parameter to construct a family of trial states for our
ground state calculations. The explicit form of the wavefunction can
be easily obtained, and is given by Eq.\ (5) of Otsuka\cite{Otsuka}
(his parameter $V$ is the same as ours). This wavefunction takes the
form of a hybridized band state, but after Gutzwiller projection, it
can also be written in the form of a overlapping Kondo cloud
state\cite{shiba}. 

 We will actually find it more convenient to
present our ground state results as a function of
\begin{equation}
  b \equiv \langle \psi_V | f^{\dagger}_{i \sigma}
  c_{i\sigma}|\psi_V\rangle.
\label{beq}
\end{equation}
Note that the average is computed {\em before} Gutzwiller projection --
in the Gutzwiller projected state the average is obviously zero. 
In the mean-field approximation, the self-consistency condition for
the homogeneous ground state reads $Jb/2\!=\!V$; this relation can
easily be worked out in the thermodynamics limit, but as stated before,
we will not use this.

We also use the Gutzwiller-projected mean field solution as our trial
state for the lowest energy triplet state in section V. The mean-field
solution in this case is inhomogeneous\cite{Wang}; hence, in this case
the parameters $V_i$ in the selfconsistency conditions $ J/2 \langle
\psi_V | f^{\dagger}_{i \sigma} c_{i\sigma}|\psi_V\rangle \!=\! V_i$
do depend on the spatial index $i$. We refer to the paper by Wang {\em
  et al.}\cite{Wang} for a detailed discussion of the structure of
this mean field solution.

The single-particle levels or our trial wave are represented by an
index for the energy the level, an index for the site, and an index
which indicates whether an electron has $c$ or $f$ character.  This
way of representing the trial wave function is suitable for the order
in which the operators in the interaction term appear in
Eq.(\ref{Ham}), and for the decoupling we have chosen to generate the
trial state. The operators between parentheses in the spin interaction
term in ${\cal H}_{KLM}$ represent intermediate steps in a Monte Carlo
diffusion process.  These correspond to changes within one spin
sector. It is, therefore, natural to have numbered electrons of a
certain spin and to allow changes from $c$ to $f$ and vice versa,
rather than to have numbered electrons with the $c$ or $f$ character
fixed and letting the spin change. Both representations are
equivalent, but our choice allows to work with Slater-matrices of
fixed size.

In the Monte Carlo calculation, the trial wave determines the
distribution of random walkers; each walker represents a
configuration, i.e., specifies the positions of each electron, its
spin, and whether it is $c$ or $f$. The weight of a certain
configuration in the initial ensemble can be calculated from the trial
state, by taking the product of the determinants corresponding to the
spin-up and spin-down single-particle states.  The ensemble is chosen
by generating configurations at random, and then comparing the weight
squared with a random number, in order to decide whether that
configuration should be accepted as a member of the ensemble or not.
By imposing the constraint of one $f$-electron per site, the ensemble
is automatically Gutzwiller-projected. Once the initial ensemble is
Gutzwiller projected, the ensemble remains so during the projection
process, since all moves allowed by ${\cal H}_{KLM}$ keep the
$f$-levels singly occupied.

\subsection{Implementation of the FNMC for the KLM}

For the KLM Hamiltonian Eq.(\ref{Ham}), all off-diagonal terms
$\langle R|{\cal H}|R\rangle$ are negative, for an antiferromagnetic
spin-interaction $J\!>\!0$.  All steps are therefore according to
(\ref{condition}) subdivided into allowed steps for which
$\psi_{\text{T}}(R)\psi_{\text{T}}(R^\prime) \!>\! 0$ and forbidden
steps for which $\psi_{\text{T}}(R)\psi_{\text{T}}(R^\prime) \!<\! 0$;
the latter steps contribute to the sign flip potential\cite{note2}
(\ref{Vsf}).  If we use ${\cal H}_{KLM}$ in the projector
(\ref{projector}), three things can happen in one time step of the
FNMC.  First of all, $R^{\prime}$ can go to a configuration with one
$c$-electron 
hopped to a neighboring position due to the kinetic term in ${\cal
  H}_{KLM}$. The second possibility is a simultaneous spin flip: the
spin interaction term proportional to $J$ allows a configuration which
has, on a certain site, a pair $c$-up, $f$-down, to change into
$c$-down, $f$-up (or vice versa). The third possibility is that
nothing happens in a time step: $R^{\prime} \!=\!R^{\prime}$. The
relative probabilities are given by Eq. (\ref{Green}). For a given
walker, which corresponds to a given configuration, a list is
therefore made of all possible allowed steps and their probabilies.
When a forbidden step is encountered in making this list, the
corresponding contribution to the sign flip potential (\ref{Vsf}) is
calculated.  Since for every configuration $R$ at most one electron
changes its state (the site- or $c/f$ label) per spin sector, the
ratio $\psi_T(R)/\psi_T(R')$ which determines the probabilities and
$V_{\text{sf}}$, can be calculated in a number of operations that is
linear in the size of the system, if one already has the transposed
inverse of the Slater matrices available\cite{note3}.

Once an ensemble of random walkers with weight determined by the trial
wave function has been prepared, as described in the previous
subsection, the Monte Carlo projection is done according to
(\ref{prod}). For a given walker, all possible moves are considered,
and for each move, the ratio's
$\psi_{\text{T}}(R)/\psi_{\text{T}}(R^\prime)$ are calculated. This
operation corresponds to a dot-product\cite{note3}, so the time needed
to compute it is linear in the system size. If
$\psi_{\text{T}}(R)/\psi_{\text{T}}(R^\prime)$ is positive, the step
to $R$ is allowed. The probability factor $G(R,R')\!=\!P(R,R')m(R')$
of allowed Monte Carlo moves are stored in an ordered table in which
each element is the sum of the previous element and the probability
factor $\tau|\psi_{\text{T}}(R)/\psi_{\text{T}}(R^\prime)|$ for a hop
or $J\tau|\psi_{\text{T}}(R)/\psi_{\text{T}}(R^\prime)|/2$ for a spin
flip.  The last element is the sum of the one but last element and the
probability factor for staying,
$G(R,R')\!=\!1-\tau(U_{\text{pot}}+V_{\text{sf}}-w)$, where
$U_{\text{pot}}+V_{\text{sf}}$ is the total {\em potential} energy of
${\cal H}_{eff}$. In this way, the value of the last element equals
$\sum_{R}G(R,R')\!=\!m(R')$; the random decision to select a move or
to stay is then made by deciding between which elements of the ordered
table the product of a random number between 0 and 1 and $m(R')$
falls, using the Numerical Recipes routine {\em locate}\cite{numrec}.

The first stage of the diffusion implementation of the projection is a
thermalization. During this stage, the parameter $w$ in
(\ref{projector}) ${\cal F}=1-\tau({\cal H}-w)$ is adjusted in such a
way that the ensemble of walkers stabilizes under the branching
process by which the multiplicities of the walkers are updated; $w$
approaches the measured ground state energy in this process.  After
the thermalization, the usual quantity being measured in a Green's
function Monte Carlo is the {\em mixed estimator}
$\langle\psi_{\text{T}}|{\cal O}|\psi^n\rangle $ for the {\em local
  value} of an operator ${\cal O}$,
\begin{equation}
  O_{\text{local}}(R)=\langle R |{\cal O}|\psi_{\text{T}}\rangle
  /\psi_{\text{T}}(R)= \left[ \sum_{R^{\prime}}\langle R |{\cal
    O}|R^{\prime}\rangle \langle R^{\prime}|\psi_{\text{T}}\rangle
\right] /\psi_{\text{T}}(R).
\end{equation}
The mixed-estimator is directly measured in the FNMC program, but a
better estimate for an expectation value is obtained by assuming that
the trial state is close to the ground state and neglecting quadratic
terms in the difference~\cite{Trivedi}:
\begin{equation}
  \langle \psi^n|{\cal O}|\psi^n \rangle \approx 2\cdot\langle
  \psi_{\text{T}}|{\cal O}|\psi^n \rangle - \langle
  \psi_{\text{T}}|{\cal O}|\psi_{\text{T}} \rangle . \label{best}
\end{equation}
We will refer to the right hand side as the {\em best estimate}.

Operators which are diagonal, like the $S_z$ spin correlation function
are most efficiently calculated, since for off-diagonal terms
computation of the ratio's
$\psi_{\text{T}}(R^{\prime})/\psi_{\text{T}}(R)$ takes substantial
computer time.  In our FNMC, $\tau$ needs to be small enough that
${\cal F}^n$ projects onto the ground state. and large enough that the
convergence is sufficiently rapid\cite{an}. We typically work with an
ensemble of on average $N_{\text{ens}}\!=\!1000$ walkers. In one {\em
  interval\/}, all walkers are propagated during $N_{\text{time}}$
time steps before branching. $N_{\text{time}}$ is chosen such that the
multiplicity factors $m$ remain less than 2.  After a thermalization
of $N_{\text{therm}}$ intervals, statistics is accumulated in
$N_{\text{block}}$ blocks of $N_{\text{intv}}$ intervals each.  In
principle, the blocks are treated as independent measurements, and
occasionally we check whether these are sufficiently independent
indeed. If necessary, we increase $N_{\text{intv}}$ to make them more
independent; an example of this will be discussed in section V. The
values of all these parameters used in the simulations will be listed
in the figure captions.

\section{Results for $J=0.2$ and $J=1.0$}\label{results}

As a first test of the FNMC on the KLM we compare with exact
diagonalization results by Yamamoto and Ueda\cite{Ueda}.  The coupling
constant is $J=0.2$, and the system consists of six sites with
periodic boundary conditions.  The trial wave functions we use are as
described in section IV.B.

In Fig.\ 2 we plot the FNMC energy as a function of $b$, defined in
(\ref{beq}). Note that the energy estimates are above the exact ground
state energy, as they should be\cite{haaf}. Furthermore, we see that
the minimum is quite flat in the range $0.15\!\lesssim \!b\! \lesssim
\!0.7$, and very close to the exact value (note the vertical scale!).
Thus, our estimate for the ground state energy one is quite
independent of the trial wave function --- apparently, therefore,
while the variational energies do depend stronly on $b$, the projected
energies are not very much affected by the fixed-node constraints over
some range of values of $b$. Finally, also note that the statistical
fluctuations are smaller close to the optimal value of $b$,
$b\!\approx\!0.25$, in agreement with the general trend that
fluctuations are smaller the better the ground state is approximated,
and that statistical fluctuations are reduced if there is a gap in the
excitation spectrum\cite{statistical} (the 1$d$ KLM does have a gap).
\begin{figure}
\hspace*{2.5cm}
\epsfysize=10cm
\epsffile{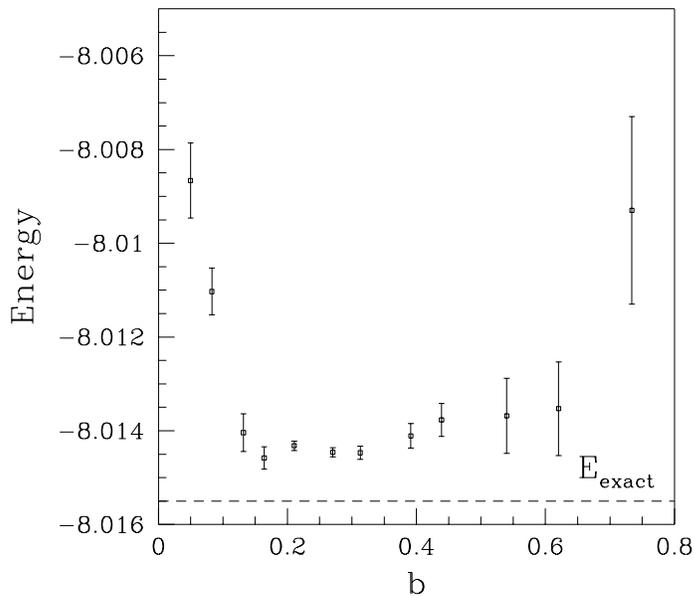}

\vspace*{0.2cm}

\caption[]{\footnotesize Energy of a six-site KLM with $J=0.2$; the parameters used in the Monte Carlo calculation are $\tau=0.01$, $w_{\text{start}}=-8.3$,
$N_{\text{time}}=30$, 
$N_{\text{therm}}=20$, $N_{\text{intv}}=5$, $N_{\text{block}}=400$,
 $N_{\text{ens}}=1000$. The dashed horizontal line denotes the exact
 result of Yamamoto and Ueda\cite{Ueda}.}
\end{figure}

\begin{figure}
\vspace{-1cm}
\epsfysize=7.5cm
\hspace{0cm}
\epsffile{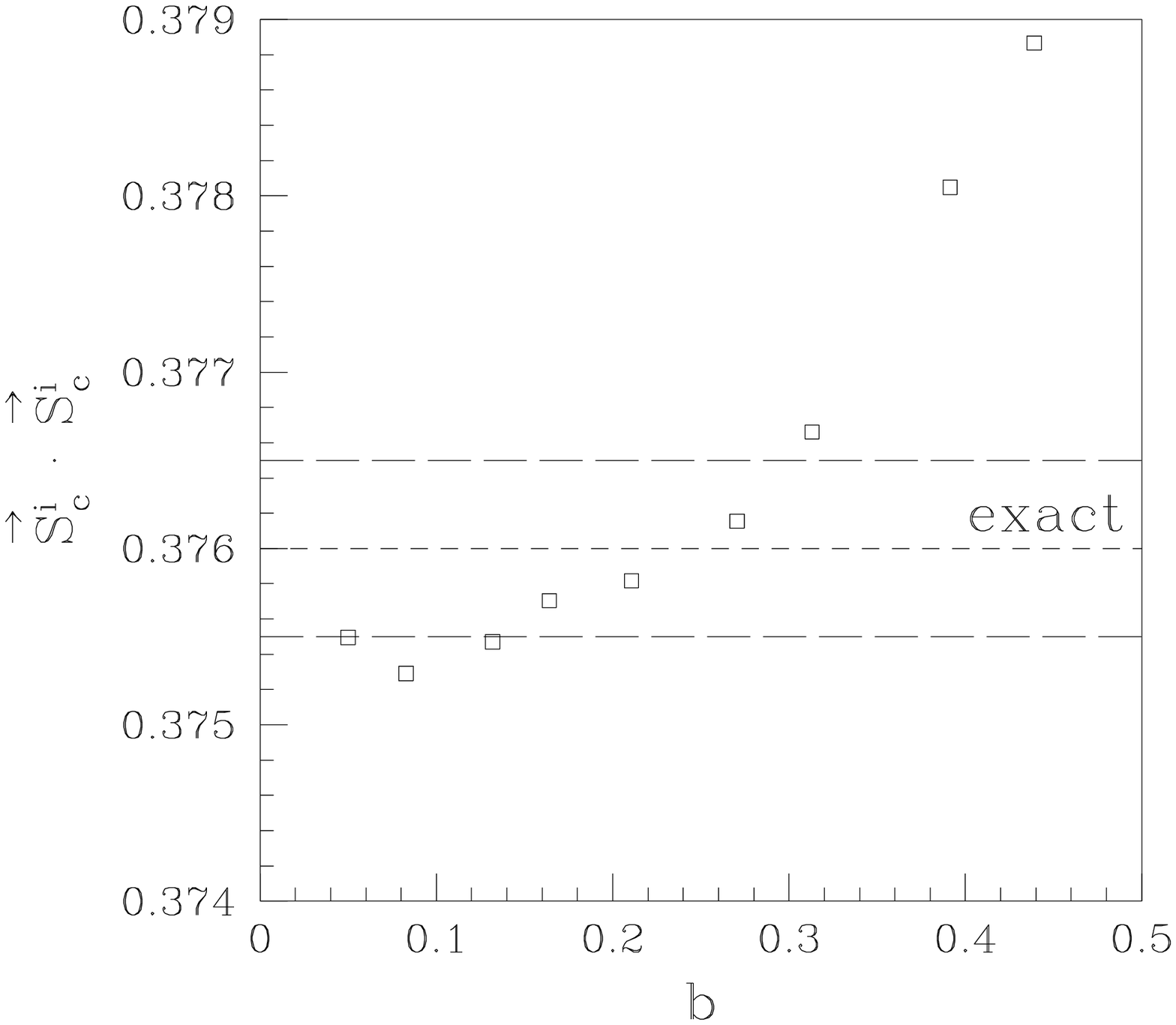}
\epsfysize=7.5cm
\hspace{0cm}
\epsffile{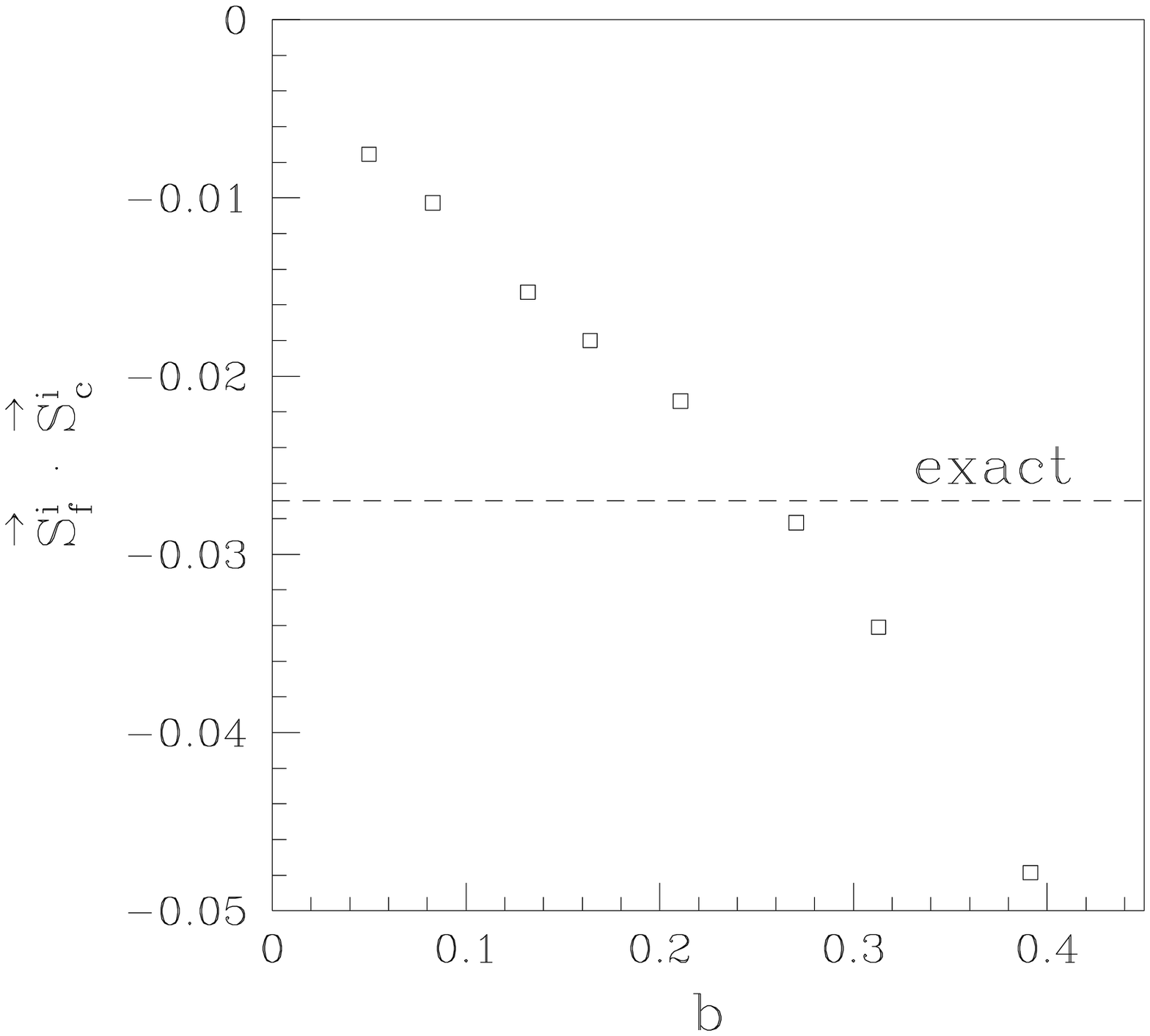}

\vspace*{1.5cm}

\caption[]{\footnotesize Two examples of correlation functions in the $J=0.2$ KLM. The parameters used in the FNMC program are
 $\tau=0.03$, $w_{\text{start}}=-8.3$,
$N_{\text{time}}=30$, 
$N_{\text{therm}}=20$, $N_{\text{intv}}=5$, $N_{\text{block}}=200$,
 $N_{\text{ens}}=1000$. Only  the mixed
 estimators are shown. The short dashed line indicates the exact value
 from Ref. \cite{Ueda}, and the long dashed lines indicate the
 precision to which this exact result was given. }

\end{figure}
An example of the mixed FNMC estimates of two spin correlation
function, $\langle \psi_T| \vec{S}^i_c \cdot
\vec{S}^i_c|\psi^n\rangle$ and $\langle \psi_T| \vec{S}^i_f \cdot
\vec{S}^{i}_c|\psi^n\rangle$ for $J\!=\!0.2$, are shown in Fig.\ 3. We
see that near the optimal value of $b$, the mixed estimate is close to
the exact value. At the same time, this figure illustrates that, not
unexpectedly, the correlation functions are more sensitively dependent
on the trial wave function than the projected energy shown in Fig.\ 2:
at $b\!\approx\!0.4$, the mixed estimate of the nearest neigbor
$c$-$f$ spin correlation function is almost a factor of 2 off, while
the energy at this value is still quite close to the proper value. As
we shall illustrate in detail below for $J=1$, the estimates improve
when we 
consider the best estimates instead.

For $J\!=\!1.0$ we follow the same procedure as for $J\!=\!0.2$ and
compare with exact results obtained by Otsuka\cite{Otsuka}.  The upper
line in Fig.\ 4a gives the energy measured in the {\em starting}
ensemble, the lower line is the Fixed-Node value, i.e., after
projection.  Like in the case of $J=0.2$, the latter curve is very
flat, while the starting values depend strongly on the input wave
function.  Fig.\ 4b shows the same data on an expanded scale, on which
one can see that the flat part of Fig.\ 4a really has a minimum.  The
exact diagonalization result\cite{Otsuka} $E=-8.561616$ is also
indicated in the picture. Clearly, also in this case the FNMC
projection method is able to come quite close to the exact energy even
if 
we start from a trial state that has a bad energy, and the statistical
fluctuations are again smallest close to the minimum in FNMC energy.
\begin{figure}

\vspace{0cm}
\epsfysize=7.5cm
\hspace{0cm}
\epsffile{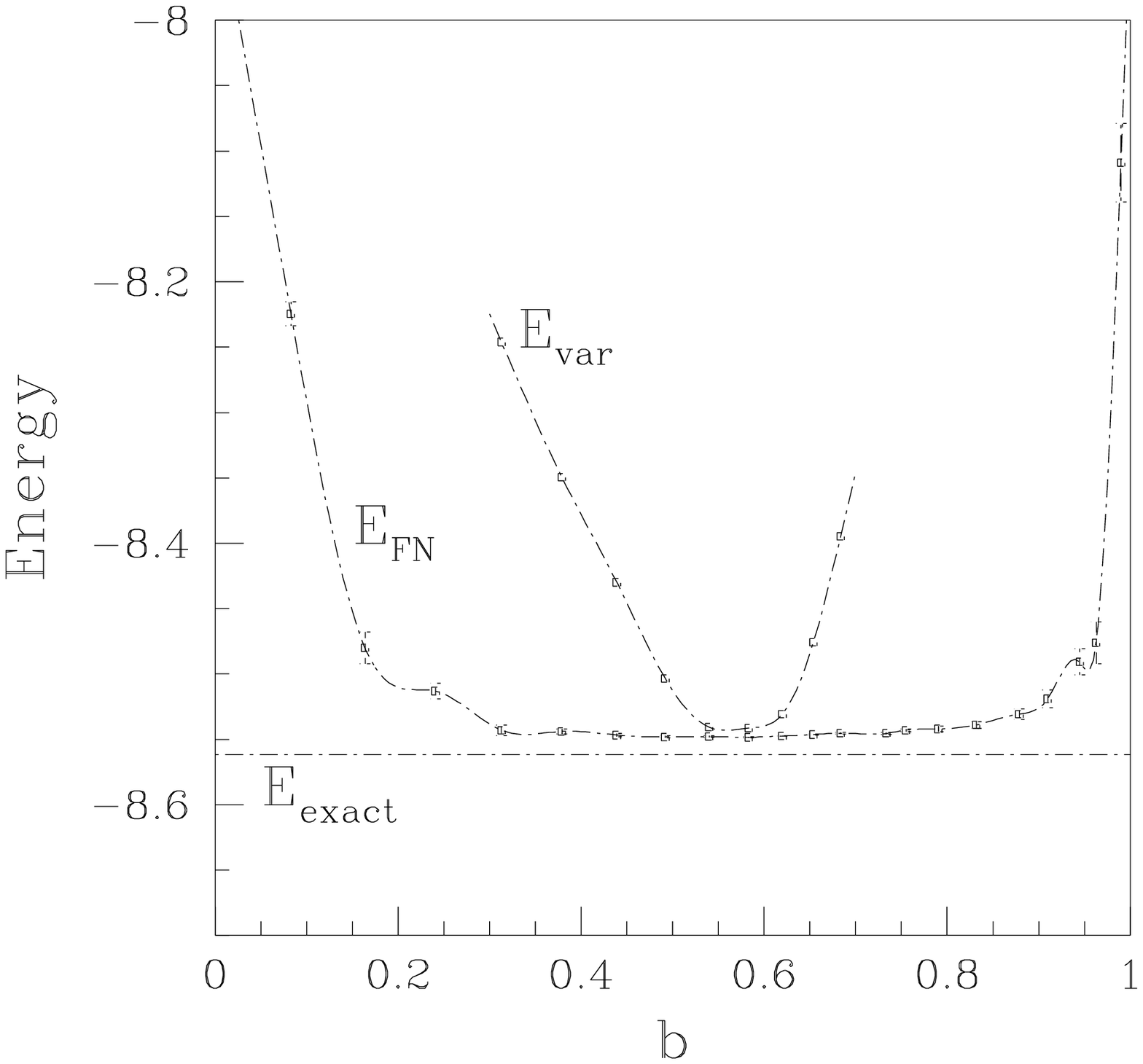}
\epsfysize=7.5cm
\hspace{0cm}
\epsffile{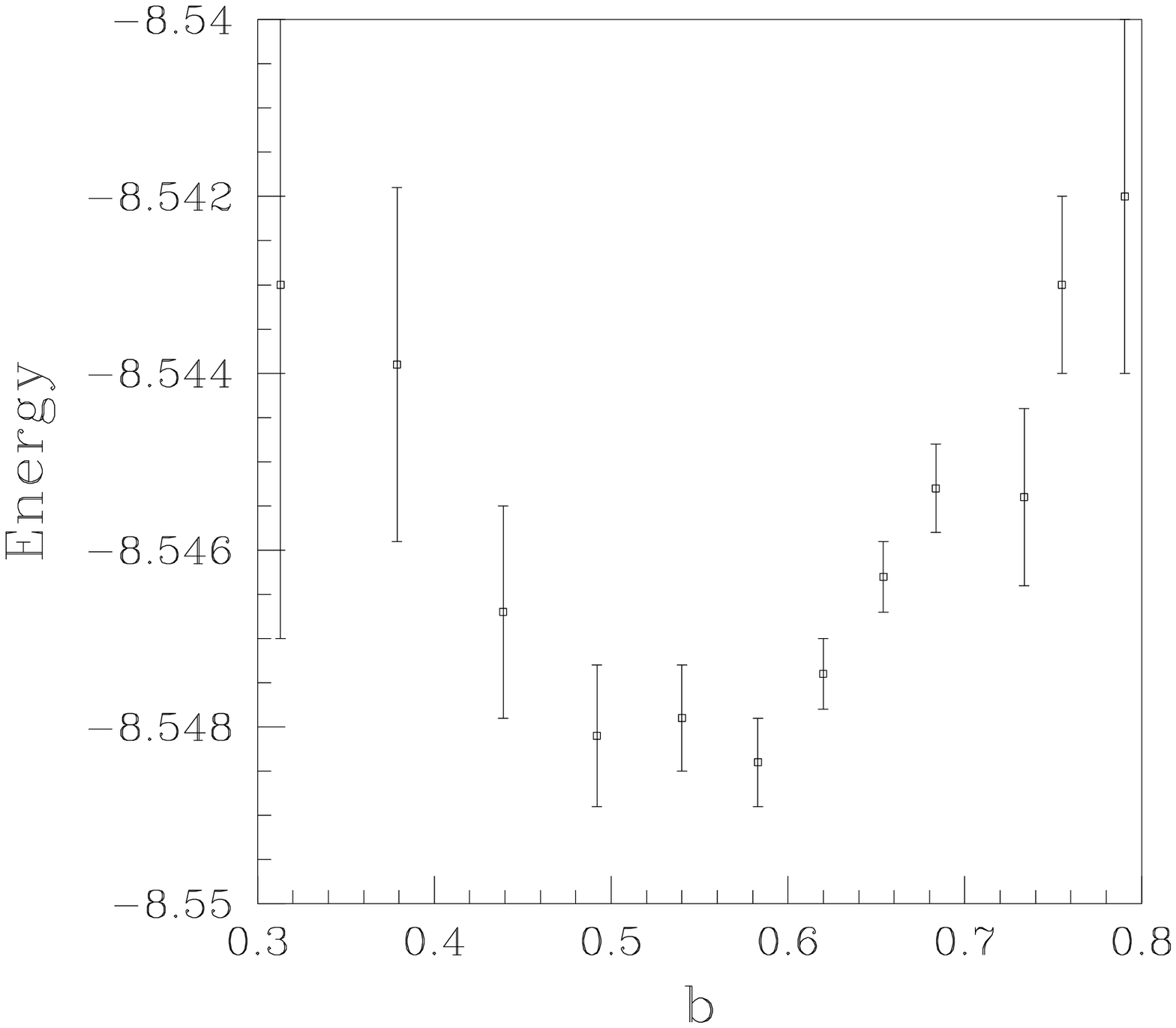}

\vspace*{1.5cm}

\caption{\footnotesize  Energy of a six-site KLM with $J=1.0$; the parameters used in
the FNMC  calculation are $\tau=0.003$, $w_{\text{start}}=-9.9$,
$N_{\text{time}}=20$, 
$N_{\text{therm}}=20$, $N_{\text{intv}}=1$, $N_{\text{block}}=10000$,
 $N_{\text{ens}}=1000$. In the right plot, the same results are
 plotted on an expanded scale. }

\end{figure}

In Figs.\ 5 and 6, we present the results for on-site correlation
functions and correlation functions involving different sites,
respectively. Three values are plotted: the variational value (using
the Gutzwiller projected state $|\psi_V\rangle$), the mixed estimator
and the best estimate given by (\ref{best}).  For all correlation
functions considered, the latter curve is relatively flat throughout
the whole range of $b$ values where the projected energy shown in Fig.
4 is close to the exact value.  Thus, although the estimated
correlations are slightly off, these results do show that, at least
for the KLM, correlation functions are not strongly dependent on the
trial state, and hence can be estimated relatively well with our FNMC.

\begin{figure}
\vspace{0cm}
\epsfysize=7.5cm
\hspace{0cm}
\epsffile{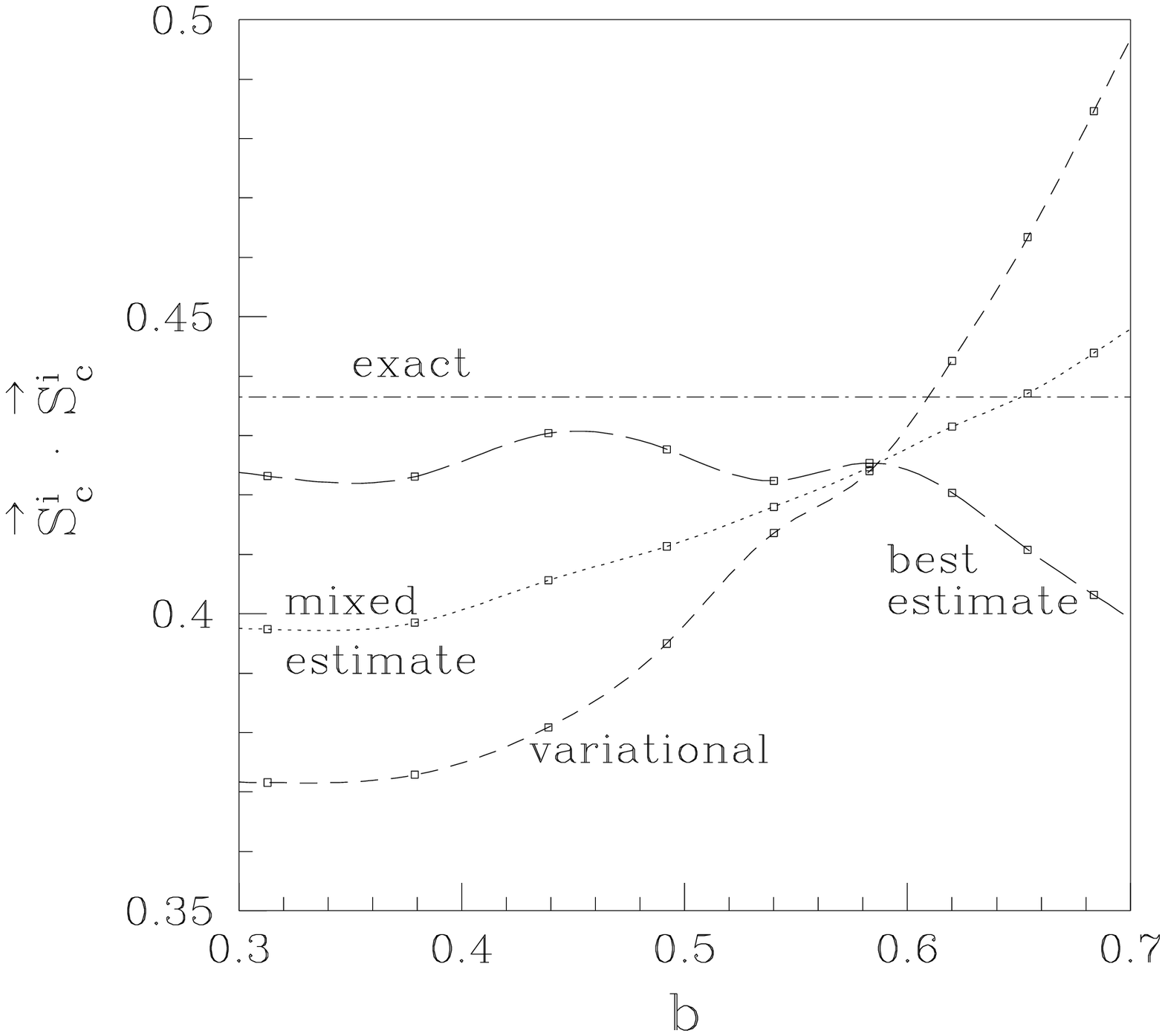}
\epsfysize=7.5cm
\hspace{0cm}
\epsffile{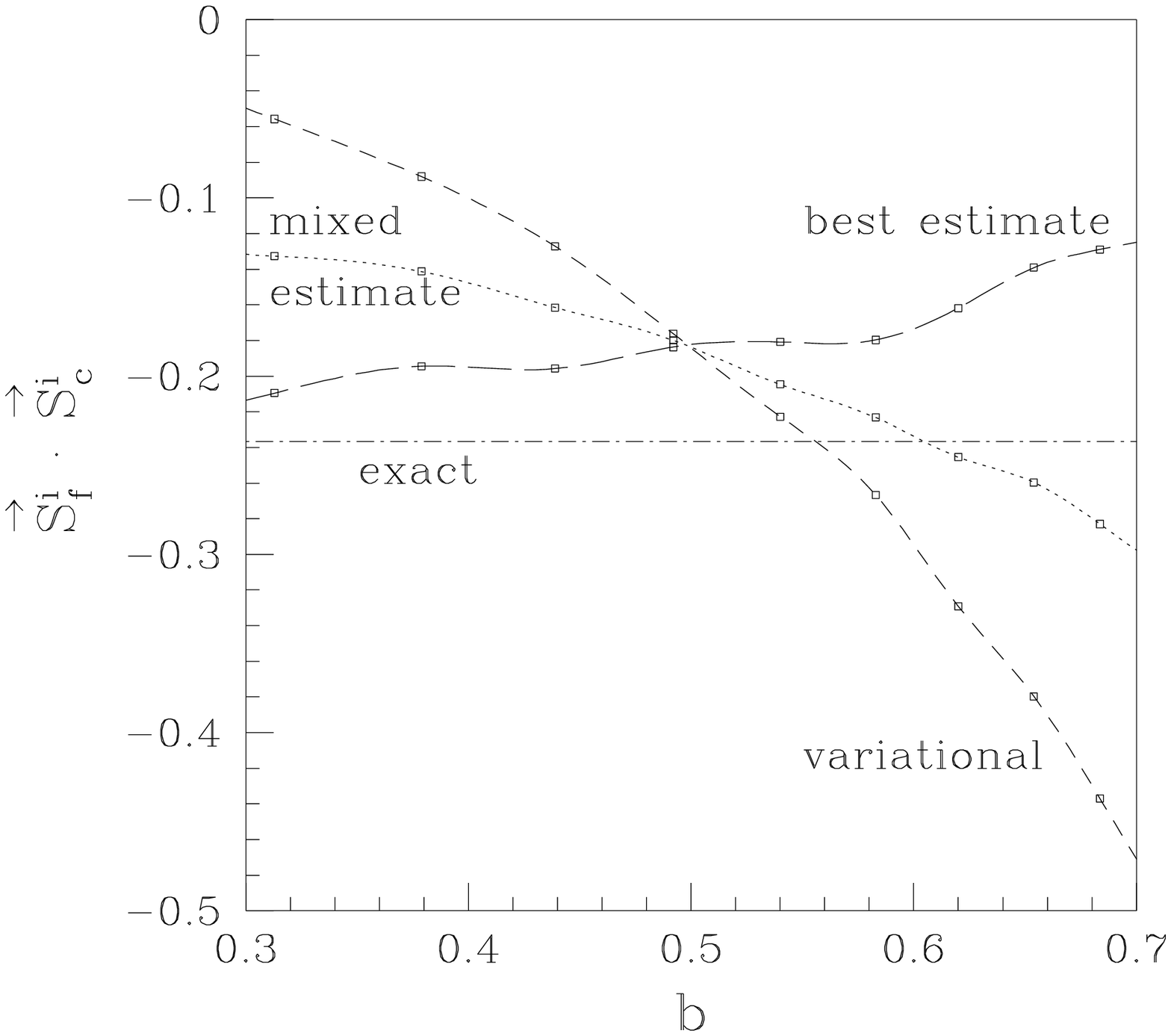}

\vspace*{1.5cm}

\caption{\footnotesize On-site correlations in the $J=1.0$ KLM.
The parameters used in the FNMC calculation are $\tau=0.003$, $w_{\text{start}}=-9.9$,
$N_{\text{time}}=20$, 
$N_{\text{therm}}=20$, $N_{\text{intv}}=1$, $N_{\text{block}}=200$,
 $N_{\text{ens}}=1000$.
 }
\end{figure}

\begin{figure}
\vspace{0cm}
\epsfysize=7.5cm
\hspace{0cm}
\epsffile{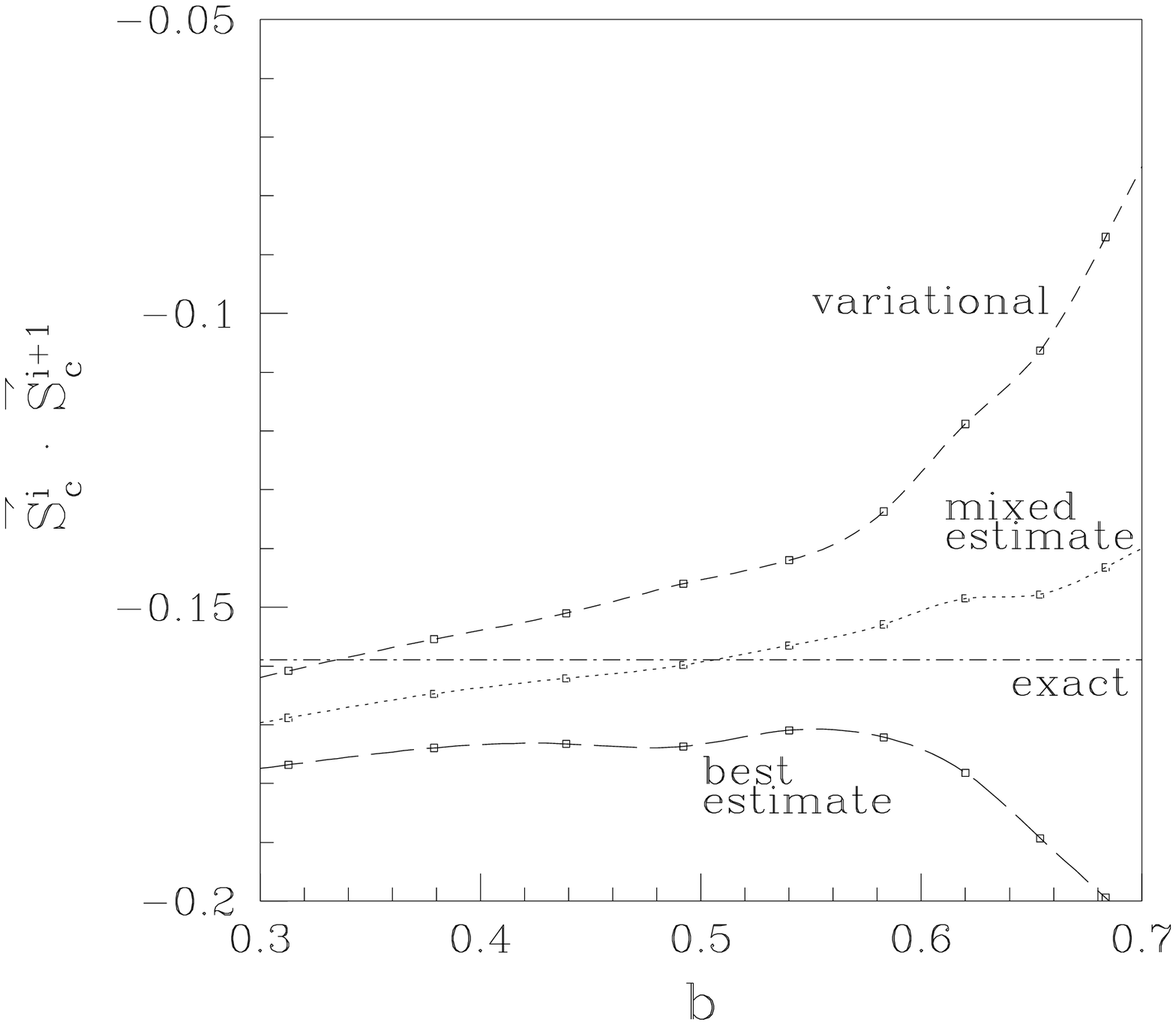}
\epsfysize=7.5cm   
\hspace{0cm}   
\epsffile{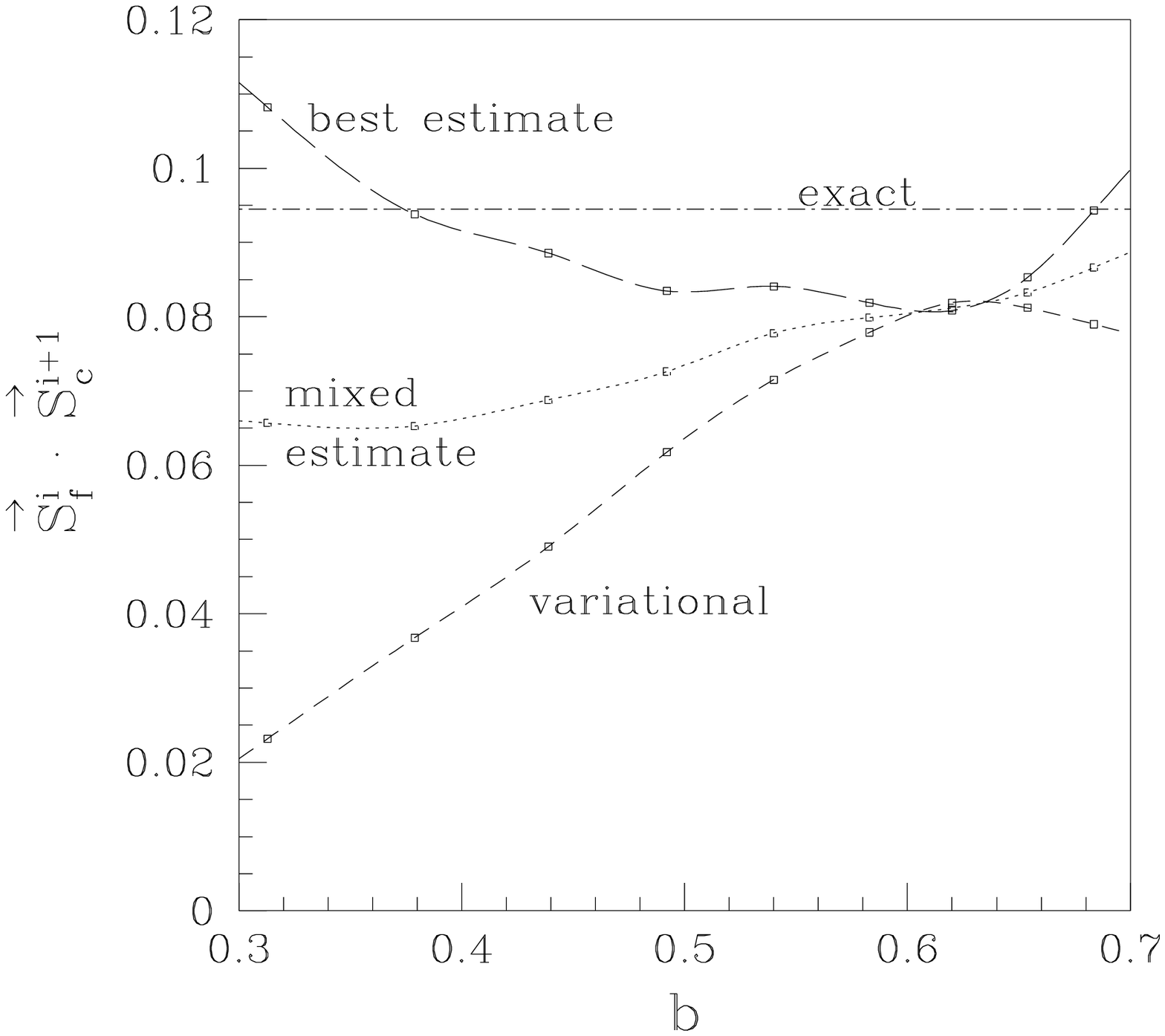}
\vspace{0cm}
\begin{center}
\epsfysize=7.5cm
\hspace{0cm}
\epsffile{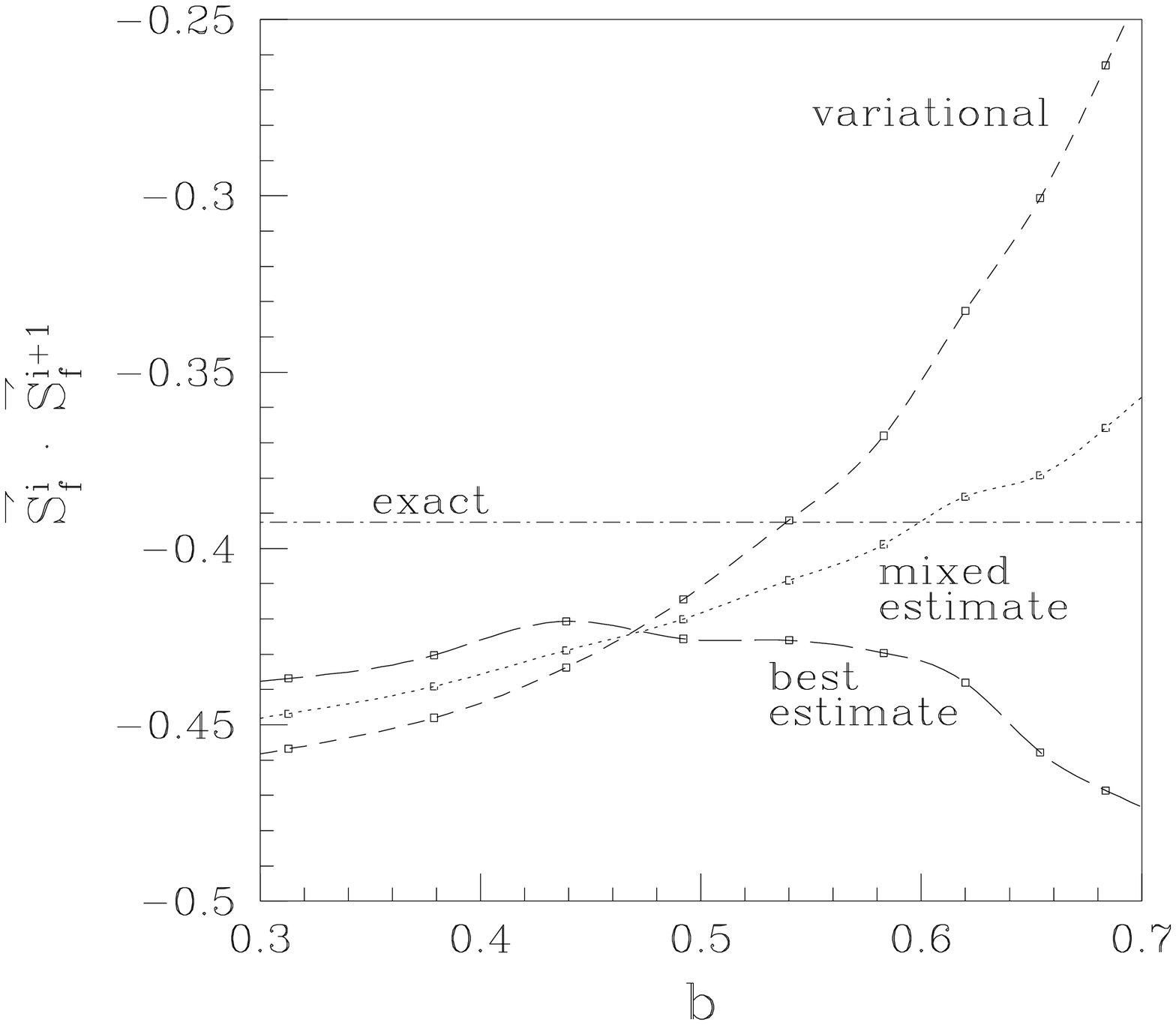}
\end{center}
\vspace*{1.5cm}

\caption{\footnotesize Correlations on different sites in the $J=1.0$ KLM.
The parameters used in the FNMC  calculation are $\tau=0.003$, $w_{\text{start}}=-9.9$,
$N_{\text{time}}=20$, 
$N_{\text{therm}}=20$, $N_{\text{intv}}=1$, $N_{\text{block}}=200$,
 $N_{\text{ens}}=1000$.
  }
\end{figure}

\section{FNMC calculation for the spin soliton}\label{soliton}

The lowest-energy excitation bove the $S=0$ ground state of the
half-filled KLM has total spin $S=1$\cite{Wang,Yu}.  In a mean-field
calculation, one is able to obtain a self-consistent solution with the
spin excitation localized on a few sites\cite{Wang}.  Wang {\em et
  al.\/}\cite{Wang} proceed by performing a Gutzwiller-projected
mean-field calculation and by writing
\begin{equation}
  |\psi_{q}\rangle=\sum_{x_{c}} \exp (i q x_{c})|\psi_{x_{c}}\rangle
  ,\label{disp}
\end{equation}
with $|\psi_{x_{c}}\rangle ={\cal P}_{G} |\psi^{mf}_{x_{c}}\rangle $
the Gutzwiller-projected local triplet state with the center of the
soliton located at $x_{c}$. The minimum of this dispersion is at wave
number $q=\pi$.  We follow the general strategy of investigating the
robustness of mean-field results by using this wave function as trial
wave function in a FNMC calculation.  To obtain the spin-gap in the
FNMC, we perform calculations both in the $S=0$ and in the $S=1$
sector.  GFMC does not always project on the ground state, only on the
lowest state that has a component along the trial-state.  Here, we use
this to our advantage: the total spin is conserved by the Hamiltonian,
and, therefore, if one starts in the $S=1$ sector, one remains in the
$S=1$ sector.  Comparing the lowest energies in both sectors gives the
gap.

Eq.\ (\ref{disp}) as it stands, seems to indicate that not only
different {\em signs} occur, but also different complex {\em phases}.
Because of reflection symmetry, however, one can combine $q$ and $-q$
and write
\begin{equation}
  |\psi_{q}\rangle=\sum_{x_{c}} \cos (q x_{c})|\psi_{x_{c}}\rangle ,
\end{equation}
which is a real problem again\cite{bemmelthesis}. 

We perform FNMC calculations for a system of 20 sites with periodic
boundary conditions. This however takes computer time: for each
possible step, 20 ratios of determinants need to be calculated, not
just one, since
\begin{equation}
  \frac{\psi_{\text{T}}(R)} {\psi_{\text{T}}(R^{\prime})}
  =\frac{\sum_{x_{c}} \cos (q x_{c})\langle R |\psi_{x_{c}}\rangle}
  {\sum_{x_{c}} \cos (q x_{c})\langle R^{\prime} |\psi_{x_{c}}\rangle}
  =\frac{ \sum_{x_{c}} \cos (q x_{c})\langle R^{\prime}
    |\psi_{x_{c}}\rangle \left( \psi_{x_{c}}(R) / 
        \psi_{x_{c}}(R')\right)
        } {\sum_{x_{c}} \cos (q x_{c})\langle R^{\prime}
        |\psi_{x_{c}}\rangle}.
\label{ratiot}
\end{equation}
  
The factors $\psi_{x_{c}}(R)/ \psi_{x_c}( R^{\prime})$ can be obtained
as simple dotproducts again\cite{note3}, in most cases.  The exception
is when $\psi_{x_{c}}(R')\!=\!0$: in those cases $\psi_{x_{c}}(R) $
needs to be calculated from scratch, for which the computer time
increases as the cube of the size of the system.

Note that this difficulty never occurs in a calculation with only one
Slater determinant as trial wave function: with importance sampling,
the probability to be in a configuration is proportional to
$\psi_{\text{T}}(R^{\prime})$, so if this is zero, a random walker
never visits such a configuration. Therefore, we never need to compute
ratio's $\psi_{\text{T}}(R)/ \psi_{\text{T}}(R^{\prime})$ in which the
{\em old} configuration $R^{\prime}$ has $\psi_{x_{c}}(R')\!=\!0$.  In
the present case, only $\sum_{x_{c}} \cos (q x_{c})\langle R^{\prime}
|\psi_{x_{c}}\rangle$ determines the probability to visit a
configuration $R^{\prime}$, and a single $\psi_{x_{c}}(R')$ may be
small or zero.

While smallness may be a practical difficulty in terms of numerical
accuracy, the main problem is that a Slater-matrix can be really
singular.  In the case of a $S=1$ soliton trial-state with $S_z=1$,
this turns out to happen for $\langle R^{\prime} |\psi_{x_{c}}\rangle$
if the $f$-electron at site $x_{c}$ has spin {\em down}, in
configuration $R^{\prime}$, as is illustrated by the results of Fig.\
2 of Wang {\em  et al.}\cite{Wang}.

Considering possible moves, $\psi_{x_{c}}(R)$ needs to be calculated
for all $x_{c}$ and for all $R$. Computation of many values
$\psi_{x_{c}}(R)$ from scratch would be very time-consuming. However,
the only possibility to make a singular matrix non-singular, is to
flip the spin of the $f$-electron (and of the $c$-electron) on site
$x_{c}$. Such a flip can make one singular matrix non-singular, and
{\em only\/} for the corresponding new configurations $R$ do we need
to calculate one Slater determinant.

In practice, we keep track of which of the 20 matrices are singular,
for a certain random walker. For all hops, all non-zero new values
$\psi_{x_{c}}(R)$ can be calculated as dotproducts, so that in
calculating the ratio (\ref{ratiot}) each flip in which one goes from
$f$-down to $f$-up can make {\em one} Slater matrix non-singular, and
for this one the determinant has to be calculated from scratch.

Once a step has been chosen, all matrices have to be updated.  If it
is a flip from $f$-up to $f$-down, one determinant becomes singular.
If it is a flip with f-down to f-up, one matrix becomes non-singular,
and its transposed inverse needs to be calculated, for facilitating
the calculation of transition probabilities of subsequent steps.
Except for this more complicated calculation of $\psi_{\text{T}}(R) /
\psi_{\text{T}}(R^{\prime})$, the FNMC program is the same as
described before.
\begin{figure}
\vspace{-1.5cm}
\epsfysize=10cm
\hspace{2cm}
\epsffile{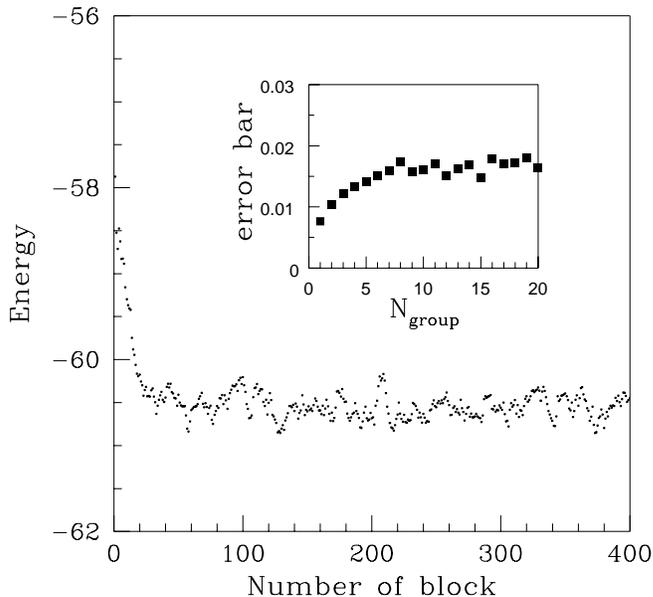}
\vspace*{1.5cm}

\caption[]{\footnotesize Illustration of how a run proceeds:
  The average Monte Carlo energy calculated during one `block' in the
  FNMC calculation for a $k=\pi$ soliton in a KLM on a 1$d$ lattice of
  20 sites with $J=4.0$, calculated with parameters $\tau = 0.01$,
  $w_{start}=-90$, $N_{therm}=3$, $N_{time}=1$, $N_{intv}=3$,
  $N_{block}=400$, $N_{ens}=200$.  The inset shows the error bar of
  the last 300 blocks, and calculated by grouping $N_{group}$
  measurements together. For $N_{group}\gtrsim 3$, the error bars do
  not increase, indicating that the energies of different groups are
  statistically independent.}
\end{figure}

Fig. 7 illustrates how the FNMC projection proceeds as the number of
iterations increases.  First, one observes projection on the ground
state: the energy measured over a number of Monte Carlo time steps, a
'block', drops.  Then, there are fluctuations around a mean value.
Accumulating statistics leads to a reduction of the error bars.  One
observes that the measurements in consecutive blocks are not
independent.  The correct error bar is obtained by grouping
$N_{\text{group}}$ measurements together and thus dividing the
$N_{\text{total}}$ blocks in $N_{\text{total}}/N_{\text{group}}$
measurements. After this regrouping, one has fewer values, but they
are more independent.  The error bars this gives are plotted in the
inset of Fig.\ 7: the value of the plateau is the error bar we report
in the energy dispersion curve (all error bars in energies are
obtained this way).
\begin{figure}
\vspace{0cm}
\epsfysize=9cm
\hspace{3cm}
\epsffile{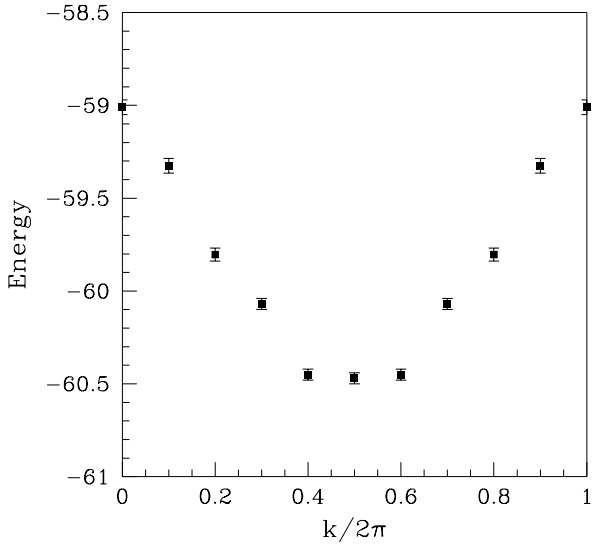}
\vspace*{1.5cm}

\caption{\footnotesize Dispersion of a running soliton
in a $J=4.0$ KLM of 20 sites. The parameters used in the Monte Carlo
 calculations are $\tau=0.005$, $w_{\text{start}}=-90$,
$N_{\text{time}}=1$, 
$N_{\text{therm}}=3$, $N_{\text{intv}}=3$, $N_{\text{block}}=200$,
 $N_{\text{ens}}=200$.
  }
\end{figure}

The FNMC dispersion of the spin soliton, for $J=4.0$, on 20 sites, is
presented in Fig.\ 8.  Since the $S=0$ value is
$E_{S=0}^{FNMC}=-63.423(5)$, and the minimum of the dispersion is at
$E_{S=1}^{FNMC}=-60.47(3)$, the gap we obtain in FNMC approximation is
$\Delta_{S}^{FNMC}=2.95(3)$, which is in agreement with the results
shown in Fig.\ 3 of
Wang {\em et al.} using a Gutzwiller-projected mean-field
approximation, and those of Yu and White\cite{Yu} using the density
matrix renormalization group method.  So, while the energies in both
the $S=0$ and the $S=1$ sector drop relative to the one estimated with
the Gutzwiller-projected mean-field wave function, the {\em
  difference} between the $S=0$ and the $S=1$ ground state energies is
essentially the same as what is known to be the correct
value\cite{Yu}.

\section{Conclusions}
We have applied the lattice FNMC method to the 1$d$ KLM. We observe
that, for small lattices, quite accurate ground state energy estimates
are obtained, even if the starting trial wave function gives  quite a
bad approximation to the energy. In the FNMC, the energy estimate is
always above the exact ground state energy, but in the cases studied
here, they are only slightly above the exact values for a large range
of values in the variational trial wave.  Over the same range over
which a good ground state energy estimate is obtained, reasonably
accurate values for correlation functions in FNMC approximation are
obtained, which again are rather independent of the starting
trial-state of the KLM. To be able to calculate a dispersion of an
excitation, one needs a conserved quantity, such that the lowest
energies in different sectors can be compared --- e.g., the isospin
gap\cite{Yu} of the KLM can not be obtained within our FNMC, but the
spin-gap can, since for its determination only ground state energies
of different spin sectors are needed.  If inversion symmetry is
present, one does not need to use complex phases and the FNMC method
for a real wave function can be applied.

In the present case, therefore, the FNMC appears to work well, in that
the constraint imposed by the fixed node condition do not appear to
have a dramatic effect on the energies and correlation functions over
a reasonably large range of values of $b$. Unfortunately, as long as
we lack more fundamental insight in the sign structure of the fermion
wavefunction, it is difficult to say whether this is just one lucky
example, or a robust property. Of course, in the present case our
trial wave function has properly built in the tendency of the $c$ and
$f$-spins to form singlets, and our fixed-node estimates are not good
in the extreme limits $b\!\rightarrow\!0$ (no local singlet
correlations) and $b\!\rightarrow \! 1$ (tightly bound singlets).

For the smaller lattice sizes we have considered here, the sign
problem does not appear to be so severe\cite{Otsuka}, but for the
larger lattice sizes, like those needed to study the spin triplet
excitation or domain walls in de two-dimensional Hubbard
model\cite{Bemmel}, the advantages of the FNMC are more prominent.

Our results also throw new light on our own earlier results for domain
walls in the two-dimensional Hubbard model\cite{Bemmel}. In these
simulations, we compared ground state energies starting from a
homogeneous trial state and from a domain wall trial state. Although
the lowest energy state was found when applying a FNMC projection to a
domain wall trial state, the ground state energy obtained after FNMC
projection of a homogeneous trial state was found to be relatively
close. Since it is conceivable that, e.g., domain wall type
correlations do build up during the projection of a homogeneous trial
state, a study of the correlation functions is needed before clear
conclusions can be drawn.

\section{Acknowledgement}
We are grateful to P.J.H. Denteneer  for
illuminating discussions and  in particular to J.M.J. van Leeuwen for
introducing us to the subject and for discussions and advice.



\begin{thebibliography}{99}
\bibitem{an2} The FNMC of Refs.  10 and 11 is based in spirit on
  the approach developed in Ref. 9, but this earlier formulation
  did not lead to an upper bound on the ground state energy.
\bibitem{misnomer}
The name `fixed-node' is actually a slight misnomer for our method,
since the nodes of a wave function are not always properly defined for
a 
discrete configuration space. In our method, the domains where the
trial wave has a given sign are held fixed.
\bibitem{note1} The sizes for which
exact-diagonalization results exist are rather small, because the size of the
Fock space is $8^{N}$, with $N$ the number of sites.
\bibitem{note2}
Because in this case the sign flip in Eq.(\ref{condition}) always corresponds
to a change in sign of the trial wave function, the sign flip
potential $V_{\text{sf}}$ is the same as the `nodal boundary potential'
used before\cite{Bemmel}; the term sign flip
potential introduced later\cite{haaf} is more widely applicable.
\end{thebibliography}
\end{document}